# Optimal design with EGM approach in conjugate natural convection with surface radiation in a two-dimensional enclosure

**Mohammad Amin Dashti[*] , Ali Safavinejad[†]**

*Department of Mechanical Engineering, University of Birjand, Birjand, Iran.*

Analysis of conjugate natural convection with surface radiation in a two-dimensional enclosure is carried out in order to search the optimal location of the heat source with entropy generation minimization (EGM) approach and conventional heat transfer parameters. The air as an incompressible fluid and transparent media is considered the fluid filling the enclosure with the steady and laminar regime. The enclosure internal surfaces are also gray, opaque and diffuse. The governing equations with stream function and vorticity formulation are solved using finite difference approach. Results include the effect of Rayleigh number and emissivity on the dimensionless average rate of entropy generation and its optimum location. The optimum location search with conventional heat transfer parameters including maximum temperature and Nusselt numbers are also examined.

[*] Email address: m.amin.dashti@birjand.ac.ir , m.amin.dashti@gmail.com

[†] Email address: asafavi@birjand.ac.ir

**Nomenclature**

| | | | |
|---|---|---|---|
| $C_p$ | specific thermal capacity of the fluid (J/kgK) | $\dot{S}_{gen,sys}$ | total rate of entropy generation for system (W/K) |
| $D_0$ | heat source strip size (m) | $\bar{\dot{S}}_{gen,sys}$ | average rate of entropy generation for system (W/m$^3$K) |
| $D_h(D_0/H)$ | dimensionless heat source strip size | $\dot{S}_{heat}$ | total rate of entropy generation due to the heat transfer (W/K) |
| $F_{kj}$ | view factor of k-th element to the j-th element of the enclosure internal surface | $\dot{S}'''_{fric}$ | volumetric rate of entropy generation due to the friction (W/m$^3$K) |
| $g$ | acceleration of the gravity (m/s$^2$) | $\dot{S}'''_{gen}$ | volumetric rate of entropy generation (W/m$^3$K) |
| $H$ | width and height of the enclosure (m) | $\dot{S}'''_{heat}$ | volumetric rate of entropy generation due to the heat transfer (W/m$^3$K) |
| $k_f$ | thermal conductivity of the fluid (W/mK) | $T$ | temperature (K) |
| $N$ | total number of internal surface elements of the enclosure | $T_C$ | temperature of cold sidewall (K) |
| $N_{rc}$ | radiation-conduction number | $u, v$ | velocity components (m/s) |
| $N_s$ | dimensionless average rate of entropy generation | $U, V$ | dimensionless velocity components |
| $N_{s,fric}$ | dimensionless average rate of entropy generation due to the friction | $x, y$ | Cartesian coordinates (m) |
| $N_{s,heat}$ | dimensionless average rate of entropy generation due to the heat transfer | $X, Y$ | dimensionless Cartesian coordinates |
| $N_{s,sys}$ | dimensionless average rate of entropy generation for system | $Y_h \left(\frac{S_0}{H}\right)$ | dimensionless heat source space from bottom |
| $N'''_s$ | dimensionless volumetric rate of entropy generation | *Greek symbols* | |
| $N'''_{s,fric}$ | dimensionless volumetric rate of entropy generation due to the friction | $\alpha$ | thermal diffusivity (m$^2$/s) |
| $N'''_{s,heat}$ | dimensionless volumetric rate of entropy generation due to the heat transfer | $\beta$ | coefficient of volumetric thermal expansion (1/K) |
| $Nu_{avg,c}$ | average convective Nusselt number | $\varphi$ | irreversibility coefficient |
| $Nu_{avg,r}$ | average radiative Nusselt number | $\delta$ | Kronecker delta |
| $Nu_{avg,tot}$ | average total Nusselt number | $\varepsilon$ | emissivity of internal surface |
| $P$ | pressure (N/m$^2$) | $\theta$ | dimensionless temperature |
| Pr | Prandtl number | $\mu$ | dynamic viscosity (kg/ms) |
| $q''$ | heat flux (W/m$^2$) | $\nu$ | kinematic viscosity (m$^2$/s) |
| $\dot{Q}$ | heat transfer rate (W) | $\rho$ | fluid density (kg/m$^3$) |
| $q''_0$ | heat source flux (W/m$^2$) | $\sigma$ | Stephan-Boltzmann constant (W/m$^2$K$^4$) |
| $q_r$ | net radiative heat flux of surface (W/m$^2$) | $\psi$ | stream function (m$^2$/s) |
| $Q_r$ | dimensionless net radiative heat flux of surface | $\Psi$ | dimensionless stream function |
| $q_{i,j}$ | irradiation of the j-th element (W/m$^2$) | $\omega$ | vorticiy (1/s) |
| $Q_{i,j}$ | dimensionless irradiation of the j-th element | $\Omega$ | dimensionless vorticiy |
| $q_{o,j}$ | radiosity of the j-th element (W/m$^2$) | *Subscripts* | |
| $Q_{o,j}$ | dimensionless radiosity of the j-th element | avg | average |
| Ra | Rayleigh number | c | convective |
| $S$ | entropy of system (J/K) | max | maximum |
| $S_0$ | heat source space from bottom (m) | mid | middle |
| $\dot{S}_{fric}$ | total rate of entropy generation due to the friction (W/K) | r | radiative |
| $\dot{S}_{gen}$ | total rate of entropy generation (W/K) | tot | total |

## 1. Introduction

Optimum design in thermal systems especially thermal management of systems has an effective importance which has created a wide number of numerical studies in this field. Some mechanical and electrical commercial programs have devoted a separate part to optimization of thermal systems. Optimum configuration is designed with conventional heat transfer parameters such as minimizing the maximum value of temperature or maximizing the Nusselt number or global thermal conductance. Entropy generation minimization through heat and fluid flow [1] in optimal design of energy systems has also been an effective approach for a better performance. This study tries to reach to an optimum design of thermal boundaries with conventional heat transfer parameters and EGM approach; along with the effect of surface radiation is also focused. It's due to the fact that several studies( such as Refs. [2-9]) have emphasized the role of radiation on natural convection. The literature view discloses innovative remarks of this study further.

Natural convection in enclosures and the effect of thermal boundaries was studied by several authors. Sharif and Mohammad [10] studied Natural convection in cavities with a constant flux heat source at the bottom wall and isothermal cooling from the sidewalls of the inclined cavity. Cheikh et al. [11] focused on the effect of the boundary condition on the natural convection due to constant flux heat source with two different size in four configuration of the boundary condition in the enclosure. Chu et al. [12] studied the effect of size and location of an isothermal heater and aspect ratio, boundary conditions on the laminar natural convection in the rectangular enclosures in order to find maximum heat transfer rate. Türkoglu and Yücel [13] investigated the location effect of the two isothermal heat source and sink on the natural convection in a square cavity. Ngo and Byon [14] studied the effect of size and location of an isothermal heat source on the natural convection in a square cavity with finite element approach. Their results included the effect of size and location of the isothermal heat source on isotherms, streamlines, and Nusselt number. Da Silva et al. [15] performed the optimal search of the constant flux heat sources distribution on a sidewall of a square enclosure with natural convection in order to minimize the maximum hot spots temperature.

Entropy generation in natural convection in two-dimensional enclosures has been studied by several authors (Refs.[16-18] and references therein). For the design of two constant flux heat sources positions in natural convection in a two-dimensional enclosure, Mukhopadhyay [19] analyzed the entropy generation. Hinojosa et al. [20] presented the effect of surface radiation on entropy generation in an open cavity with isothermal heat source.

Regarding the presented studies by Authors, concerning the studies involving the effect of heat sources on natural convection, surface radiation has been considered moderately [3-9] and heat source type is isothermal. However, in the studies around the effect of heat source location, surface radiation has not been considered [10-15]. In the current study, conjugate natural convection with surface radiation in a two-dimensional enclosure including a constant heat flux source is analyzed to study the effect of surface radiation on the rate of entropy generation. Then, the entropy generation minimization approach is utilized to search the optimal location of the heat source. The other heat transfer parameters are also studied and optimal location of constant flux heat source is searched. Along with, results are compared with pure natural convection and related studies.

## 2. Problem definition and mathematical formulation

Problem geometry and boundary condition have been shown in Fig. 1. In this configuration, the parameters $S_0$ and $D_0$ show heat source space from the bottom and heat source strip size, respectively. The flux of heat source is uniform and set fixed in all computations. The enclosure is considered filled by air with constant thermophysical properties at $T_c = 293.5$ K. The opposite wall to the heat source is the sink and kept at $T_c$ constant temperature. Other boundaries are adiabatic.

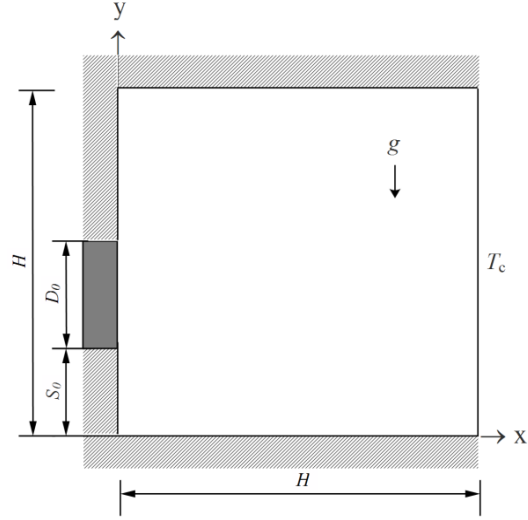

Fig. 1. The geometry and boundary conditions of the problem

## 2.1. Governing equations

The fluid flow is considered laminar, steady and two-dimensional. Air as an incompressible fluid with Boussinesq approximation and transparent media was considered the fluid filled the enclosure. So, the governing equations in primitive variables are as follow:

$$\frac{\partial u}{\partial x} + \frac{\partial v}{\partial y} = 0 \tag{1}$$

$$u\frac{\partial u}{\partial x} + v\frac{\partial u}{\partial y} = -\frac{1}{\rho}\frac{\partial P}{\partial x} + \nu\left(\frac{\partial^2 u}{\partial x^2} + \frac{\partial^2 u}{\partial y^2}\right) \tag{2}$$

$$u\frac{\partial v}{\partial x} + v\frac{\partial v}{\partial y} = -\frac{1}{\rho}\frac{\partial P}{\partial y} + \nu\left(\frac{\partial^2 v}{\partial x^2} + \frac{\partial^2 v}{\partial y^2}\right) + g\beta(T - T_c) \tag{3}$$

$$u\frac{\partial T}{\partial x} + v\frac{\partial T}{\partial y} = \left(\frac{k_f}{\rho c_p}\right)\left(\frac{\partial^2 T}{\partial x^2} + \frac{\partial^2 T}{\partial y^2}\right) \tag{4}$$

According to the thermal analysis and non-following the pressure, by differentiating from Eq. (2) with respect to $y$ and differentiating from Eq. (3) with respect to $x$ and subtracting them from each other, pressure is eliminated. By using of $u = \frac{\partial \psi}{\partial y}$, $v = -\frac{\partial \psi}{\partial x}$ and $\omega = \frac{\partial v}{\partial x} - \frac{\partial u}{\partial y}$, flow variables are replaced with stream function, vorticity, and temperature. Transform to the dimensionless form of governing and surface radiation equations (see section 2.2. Surface radiation equations) is done by dimensionless variables as follows and lead to Rayleigh, Prandtl and conduction-radiation number appear:

$$X = \frac{x}{H} \quad , Y = \frac{y}{H} \quad , \Psi = \frac{\psi Pr}{\nu} \quad , \Omega = \frac{\omega H^2 Pr}{\nu} , U = \frac{\partial \Psi}{\partial Y} \quad , V = -\frac{\partial \Psi}{\partial X} \quad , \theta = \frac{(T - T_c)}{q_0'' \frac{H}{k_f}} \quad , \theta_0 = \frac{T_c}{q_0'' \frac{H}{k_f}}$$

$$Q_r = \frac{q_r}{\sigma T_c^4} \quad , \quad Q_{o,j} = \frac{q_{o,j}}{\sigma T_c^4} \quad , \quad Q_{i,j} = \frac{q_{i,j}}{\sigma T_c^4} \quad , \quad \Pr = \frac{\nu}{\alpha} \quad , \quad \mathrm{Ra} = \frac{g\beta q_0'' H^4}{\alpha \nu k_f} \quad , \quad \mathrm{N_{rc}} = \frac{\sigma T_c^4}{q_0''} \quad (5)$$

By using dimensionless variables in Eq. (5), the dimensionless forms of governing equations are gained as follows:

$$\frac{\partial^2 \Psi}{\partial X^2} + \frac{\partial^2 \Psi}{\partial Y^2} = -\Omega \tag{6}$$

$$U \frac{\partial \Omega}{\partial X} + V \frac{\partial \Omega}{\partial Y} = \Pr \left( \frac{\partial^2 \Omega}{\partial X^2} + \frac{\partial^2 \Omega}{\partial Y^2} \right) - \mathrm{Ra} \cdot \Pr \frac{\partial \theta}{\partial X} \tag{7}$$

$$U \frac{\partial \theta}{\partial X} + V \frac{\partial \theta}{\partial Y} = \frac{\partial^2 \theta}{\partial X^2} + \frac{\partial^2 \theta}{\partial Y^2} \tag{8}$$

## 2.2. Surface radiation equations

The enclosure internal surfaces are assumed gray, opaque and diffuse. The media is also considered transparent. The net radiative heat flux of surface $q_\mathrm{r}$ by using net radiation method [21] is calculated:

$$\sum_{j=1}^{N} (\delta_{kj} - F_{kj}) q_{o,j} = q_{\mathrm{r},k} \qquad 1 \le k \le N \tag{9}$$

In Eq. (9), $N$ and $q_{o,j}$ are the total number of internal surface elements of the enclosure and radiosity of the j-th element, respectively. The view factor $F_{kj}$ are calculated by Hottel crossed string method [21]. For calculating $q_{o,j}$ in an enclosure with $N$ guessed and definite temperature elements, a system of equations is solved:

$$\sum_{j=1}^{N} [\delta_{kj} - (1 - \varepsilon_k) F_{kj}] q_{o,j} = \varepsilon_k \sigma T_k^4 \quad 1 \le k \le N \tag{10}$$

By dividing Eqs. (9) and (10) per $\sigma T_c^4$, dimensionless surface radiation equations are gained:

$$\sum_{j=1}^{N} (\delta_{kj} - F_{kj}) Q_{o,j} = Q_{\mathrm{r},k} \qquad 1 \le k \le N \tag{11}$$

$$\sum_{j=1}^{N} [\delta_{kj} - (1 - \varepsilon_k) F_{kj}] Q_{o,j} = \varepsilon_k \left( \frac{\theta}{\theta_0} + 1 \right)^4 \quad 1 \le k \le N \tag{12}$$

## 2.3. Boundary conditions

Regarding to no slip condition on the internal surface of the enclosure ($u = 0, v = 0$), it can be concluded: $\frac{\partial \psi}{\partial x} = \frac{\partial \psi}{\partial y} = 0$. The joint points of surfaces at the corners lead to $\psi = constant\ value$, and no entering or exiting the fluid into or out of the enclosure lead to $\psi = 0$ at all internal points of enclosure. Overall, dimensionless stream function and vorticity boundary conditions are written as follows:

$$Y = 0\ ,\ Y = 1: \quad \Psi = 0\ ,\quad \Omega = -\frac{\partial^2 \Psi}{\partial Y^2} \tag{13}$$

$$X = 0\ ,\ X = 1: \quad \Psi = 0\ ,\quad \Omega = -\frac{\partial^2 \Psi}{\partial X^2} \tag{14}$$

Regarding to the balance energy relation on each wall, thermal boundary conditions are attained. From the thermal boundary condition, radiation-conduction number is also achieved.

Heat source wall surface:

$x = 0$ :
$$S_0 < y < S_0 + D_0\ :\ q_c + q_r = q_0'' \quad \rightarrow \quad -k_f \frac{\partial T}{\partial x} + q_r = q_0'' \tag{15}$$

$$0 < y < S_0\ ,\ S_0 + D_0 < y < H:\ q_c + q_r = 0 \quad \rightarrow \quad -k_f \frac{\partial T}{\partial x} + q_r = 0 \tag{16}$$

Right wall surface:

$$x = H\ :\ T = T_c \tag{17}$$

Isolated wall surfaces:

$$y = 0\ :\ q_c + q_r = 0 \quad \rightarrow \quad -k_f \frac{\partial T}{\partial y} + q_r = 0 \tag{18}$$

$$y = H\ :\ k_f \frac{\partial T}{\partial y} + q_r = 0 \tag{19}$$

The dimensionless thermal boundary conditions equations are derived by using the terms in Eq. (5):

$X = 0$
$$\frac{S_0}{H} < Y < \frac{S_0 + D_0}{H} \qquad\qquad\qquad :\ -\frac{\partial \theta}{\partial X} + N_{rc} Q_r = 1 \tag{20}$$

$$0 < Y < \frac{S_0}{H}\ ,\quad \frac{S_0 + D_0}{H} < Y < 1 \quad :\ -\frac{\partial \theta}{\partial X} + N_{rc} Q_r = 0 \tag{21}$$

$$X = 1\quad :\ \theta = 0 \tag{22}$$

$$Y = 0\quad :\ -\frac{\partial \theta}{\partial Y} + N_{rc} Q_r = 0 \tag{23}$$

$$Y = 1\quad :\ \frac{\partial \theta}{\partial Y} + N_{rc} Q_r = 0 \tag{24}$$

## 2.4. Entropy generation and conventional heat transfer parameters

From the standpoint of thermodynamics second law, entropy generation can be caused due to the irreversibility of the heat transfer or viscous losses irreversibility. However, for a closed system or control mass, entropy generation around the system boundary is calculated as follows:

$$\dot{S}_{gen,sys} = \frac{dS}{dt} - \oint \frac{\delta \dot{Q}}{T} \tag{25}$$

The Eq. (25) indicates that the heat transfer by limited temperature difference is always associated with irreversibility. Regarding that the heat transfer enters the system only from the heat source boundary and deserts the system only from the cold sidewall and steady state analysis of the problem ($\Delta S = 0$), Eq. (25) is changed to:

$$\dot{S}_{\text{gen,sys}} = -\int_{S_0}^{S_0+D_0} \frac{q_0''}{T_s} dy + \int_0^H \frac{q''}{T_c} dy \tag{26}$$

Subsequently, by taking into account a depth of unit for enclosure, the rate of entropy generation is averaged over the volume of enclosure and transformed in dimensionless form regarding the system as a control mass:

$$\bar{\dot{S}}_{\text{gen,sys}} = \frac{\dot{S}_{\text{gen,sys}}}{H^2} \tag{27}$$

$$N_{\text{s,sys}} = \bar{\dot{S}}_{\text{gen,sys}} \frac{H^2}{k_f} \tag{28}$$

Furthermore, local volumetric rate of entropy generation is defined according to [1]:

$$\dot{S}_{\text{gen}}''' = \dot{S}_{\text{heat}}''' + \dot{S}_{\text{fric}}''' \tag{29}$$

Corresponding to Eq. (29), the first term shows the irreversibility due to the heat transfer in the direction of finite temperature gradients and the second term is related to the friction contribution respectively:

$$\dot{S}_{\text{heat}}''' = \frac{k_f}{T^2}\left[\left(\frac{\partial T}{\partial x}\right)^2 + \left(\frac{\partial T}{\partial y}\right)^2\right] \tag{30}$$

$$\dot{S}_{\text{fric}}''' = \frac{\mu}{T}\left\{2\left[\left(\frac{\partial u}{\partial x}\right)^2 + \left(\frac{\partial v}{\partial x}\right)^2\right] + \left(\frac{\partial u}{\partial y} + \frac{\partial v}{\partial x}\right)^2\right\} \tag{31}$$

By integrating from the Eq. (29), the total rate of entropy generation is defined as follows that should be equal to $\dot{S}_{\text{gen,sys}}$ for pure natural convection cases:

$$\dot{S}_{\text{heat}} = \int_v \dot{S}_{\text{heat}}''' \, dv \tag{32}$$

$$\dot{S}_{\text{fric}} = \int_v \dot{S}_{\text{fric}}''' \, dv \tag{33}$$

$$\dot{S}_{\text{gen}} = \dot{S}_{\text{heat}} + \dot{S}_{\text{fric}} \tag{34}$$

The dimensionless volumetric rate of entropy generation is defined:

$$N_{\text{s,heat}}''' = \dot{S}_{\text{heat}}''' \frac{H^2}{k_f}, \quad N_{\text{s,fric}}''' = \dot{S}_{\text{fric}}''' \frac{H^2}{k_f}, \quad N_s''' = \dot{S}_{\text{gen}}''' \frac{H^2}{k_f} \tag{35}$$

By integrating from the Eq. (29), the average rate of entropy generation is defined as follows in dimensionless form:

$$N_{\text{s,heat}} = \frac{1}{v}\int_v N_{\text{s,heat}}''' \, dv \tag{36}$$

$$N_{\text{s,fric}} = \frac{1}{v}\int_v N_{\text{s,fric}}''' \, dv \tag{37}$$

Thus, the dimensionless average rate of entropy generation due to the contribution of heat transfer and friction are gained:

$$N_{s,\text{heat}} = \int_0^1 \int_0^1 \frac{1}{(\theta+\theta_0)^2}\left[\left(\frac{\partial\theta}{\partial X}\right)^2 + \left(\frac{\partial\theta}{\partial Y}\right)^2\right] dXdY \qquad (38)$$

$$N_{s,\text{fric}} = \int_0^1 \int_0^1 \frac{\varphi}{\theta+\theta_0}\left\{2\left[\left(\frac{\partial U}{\partial X}\right)^2 + \left(\frac{\partial V}{\partial X}\right)^2\right] + \left(\frac{\partial U}{\partial Y} + \frac{\partial V}{\partial Y}\right)^2\right\} dXdY \qquad (39)$$

$$N_s = N_{s,\text{heat}} + N_{s,\text{fric}} \qquad (40)$$

Where $\varphi = \frac{\alpha^2 \mu}{q_0'' H^3}$ is considered the irreversibility coefficient.

Another important parameter in the optimal design of heat source position is the maximum temperature which occurs on the heat source wall in the current configuration and can be regarded as objective to minimize in dimensionless form:

$$\theta_{\max} = \frac{(T_{\max} - T_c)}{q_0'' \frac{H}{k_f}} \qquad (41)$$

Average convective, radiative and total Nusselt numbers on the heat source surface that is a constant flux heat source are defined as follows:

$$\text{Nu}_{\text{avg},c} = \frac{-1}{\frac{D_0}{H}} \int_{\frac{S_0}{H}}^{\frac{S_0+D_0}{H}} \frac{1}{\theta} \frac{\partial\theta}{\partial X} dY \qquad (42)$$

$$\text{Nu}_{\text{avg},r} = \frac{1}{\frac{D_0}{H}} \int_{\frac{S_0}{H}}^{\frac{S_0+D_0}{H}} \frac{1}{\theta} N_{rc} Q_r \, dY \qquad (43)$$

$$\text{Nu}_{\text{avg,tot}} = \text{Nu}_{\text{avg},r} + \text{Nu}_{\text{avg},c} \qquad (44)$$

## 3. Method of solution

### 3.1. Numerical method

The governing Eqs. (6), (7) and (8) are discretized in a uniform grid by finite difference method. The second order central difference scheme is used for discretizing the convective and diffusion terms. For discretizing the first derivate in boundary conditions, second order backward or forward difference scheme is used (depending on the position at the first or end of the grid). The Poisson Eq. (6) and energy Eq. (8) are solved by alternative direction implicit (ADI) method while the vorticity transfer Eq. (7) is solved by point successive under relaxation method (PSUR) or literally Gauss-Seidel iteration method. The Numerical results show this procedure has increased the convergence speed several times faster than solving all equations by ADI method in larger Rayleigh regimes. In order to eliminate the nonlinear behavior and flow disturbances, internal iteration (5 to 10 iteration depending on the Rayleigh number) is used for solving the Eq. (6) in each general iteration of solving governing equations. Numerical experience has shown that Poisson Eq. (6) for calculating stream function that is in fact continuity equation in primitive variables behaves well when accelerating is done. That is why it is suggested to decrease error of solving Eq. (6) between two consecutive iterations more with respect to Eqs. (7) and (8). It should be noted that in the non-conservative form of governing equations, accelerating in solving vorticity and energy transfer equations such as utilizing ADI solver or successive over relaxation coefficient lead to divergence.

In natural convection combined with surface radiation, afterwards initialization is done; Eq. (6) is solved by ADI solver. Then Eq. (7) is solved by PSUR and vorticity boundary condition is updated. Energy Eq. (8) is solved by ADI solver. Subsequently, boundary elements temperatures are used in net radiation method and radiosity $Q_o$ is gained by solving the system of equations based on Eq. (12). So, the net radiative heat flux of each surface element is calculated by Eq. (11) and is used to update energy equation boundary conditions. This procedure is continued to satisfy convergence criteria.

### 3.2. Grid independence check and convergence

In order to check the effect of grid nodes number on the numerical solution, four uniform grids with 31×31, 51×51, 71×71 and 91×91 size are tested based on the solution of the natural convection combined with surface radiation with $D_h = 0.5$, $\varepsilon = 0.8$, and $Ra = 10^6$ adjustments. The results of grid size independence in

Table 1 show that relative error for the average total Nusselt number and maximum dimensionless temperature in 71×71 and 91×91 is less than 0.3 %. Furthermore, Fig. 2 shows that temperature profile solution for both grid size 71×71 and 91×91 correspond to each other, so regarding precision and evaluation time, grid size 71×71 is chosen for all results. The convergence criterion is also defined based on the following relation in which $\xi$ can be each flow variable:

$$\sum_{j=1}^{N}\sum_{i=1}^{N}\left|\frac{\xi_{i,j}^{\text{new}} - \xi_{i,j}^{\text{old}}}{\xi_{i,j}^{\text{new}}}\right| < 10^{-7} \qquad (45)$$

Table 1. Results of average total Nusselt number and maximum dimensionless temperature in grid size independency check

| grid size | $\theta_{max}$ | $Nu_{avg,tot}$ | elapsed time (s) |
|---|---|---|---|
| 31×31 | 0.14760 | 9.028 | 16.6 |
| 51×51 | 0.14309 | 8.845 | 106.4 |
| 71×71 | 0.14216 | 8.790 | 387.7 |
| 91×91 | 0.14183 | 8.774 | 1036.5 |

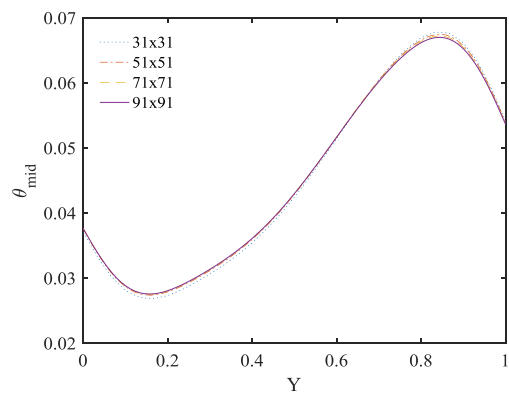

Fig. 2. Dimensionless temperature profile at midplane X=0.5 for different grid size

### 3.3. Code validation

The numerical procedure is validated in step by step way with several studies. At first, pure natural convection results based on geometry and boundary conditions in the study by Davis [22] are compared. Then, a constant heat flux source is added to pure natural convection and validation is performed based on geometry and boundary conditions in the study by Sharif and Mohammad [10]. A good agreement is shown for the Nusselt number and maximum dimensionless temperature at aspect ratio 1, inclination angle 0º, dimensionless heat source length 0.8 and different Grashof numbers in Table 2. The relative error for all quantities is less than 0.2%. A good agreement has also been gained for isotherms in the current study and study by Sharif and Mohammad [10] for aspect ratio 1, inclination angle 0º and dimensionless heat source length 0.4 in Fig. 3. Then, the validation of the natural convection conjugate with surface radiation is carried out corresponding to the geometry and boundary conditions in Wang et al. study [9]. Fig. 4 shows that isotherms and streamlines for $Ra = 10^6$ and $\varepsilon = 0.4$ in both current and Wang et al. [9] studies agree with each other.

Table 2. Validation of current study with study by Sharif and Mohammad [10] for Nusselt number (first row), maximum temperature (second row) at aspect ratio 1, inclination angle zero and dimensionless heat source length 0.8.

| Gr | current study | Sharif and Mohammad [10] |
|---|---|---|
| $10^3$ | 3.574581 | 3.55618 |
|  | 0.36536 | 0.36373 |
| $10^4$ | 3.708748 | 3.691916 |
|  | 0.369269 | 0.3674 |
| $10^5$ | 5.887121 | 5.864436 |
|  | 0.265867 | 0.26514 |
| $10^6$ | 9.307324 | 9.287972 |
|  | 0.180114 | 0.17925 |

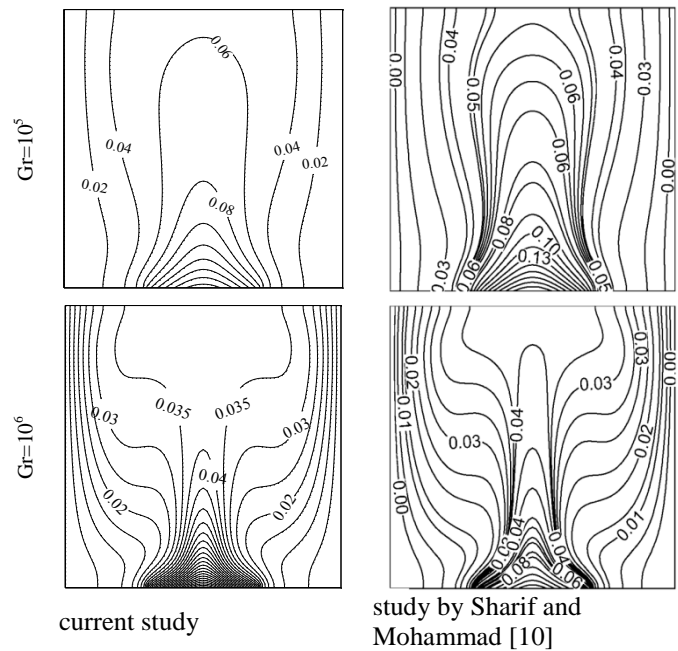

current study    study by Sharif and Mohammad [10]

Fig. 3. Dimensionless isotherms in the current study and study by Sharif and Mohammad [10] for aspect ratio 1, inclination angle 0º and dimensionless heat source length 0.4.

|   Isotherms   |   |   streamlines   |   |
|---|---|---|---|

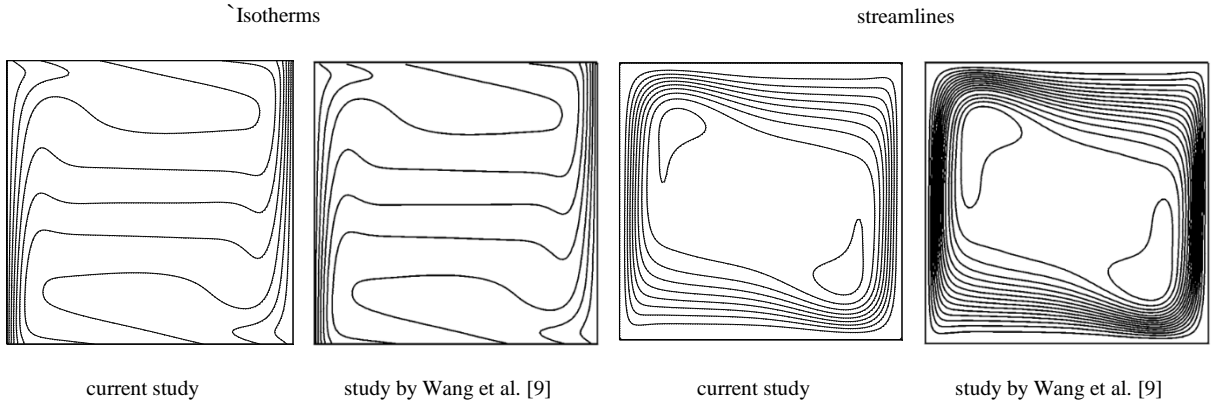

| current study | study by Wang et al. [9] | current study | study by Wang et al. [9] |

Fig. 4. Dimensionless isotherms and streamlines field in the current study and Wang et al. [9] for $Ra = 10^6$ and $\varepsilon = 0.4$

## 4. Results and discussion

### 4.1. Parameters adjustments

Analysis of conjugate natural convection with surface radiation is surveyed in a wide range of Rayleigh $10^3$ to $10^6$ and emissivity 0 to 0.8. The variations of the dimensionless average rate of entropy generation, the minimum value of the maximum dimensionless temperature of the heat source and the average Nusselt numbers with $\varepsilon$ and Ra are examined. It should be noted that regarding to the Rayleigh number definition and contribution of natural convection and surface radiation and assuming constant properties for air, Rayleigh number is a function of $q_0''$ and $H$. As discussed before, $q_0''$ is chosen fixed ($q_0'' = 10 \frac{W}{m^2}$), consequently Rayleigh number is a function of $H$ and $H$ is adjusted in relevant parameters. In fact, when surface radiation enters the natural convection the problem is not studied independently of radiation. This subject has been regarded in [7, 9].

### 4.2. Effect of surface radiation on the flow stream and temperature field

Before analysis the rate of entropy generation, it is useful to evaluate the effect of surface radiation on the flow stream and temperature field; hence firstly the velocity component profiles at the midplane of the enclosure and the temperature of heat source wall surface for $Ra = 10^3$ and $Ra = 10^6$ are presented in Fig. 5 and Fig. 6, respectively. Corresponding to Fig. 5a, for $Ra = 10^3$ as emissivity increases, both velocity components decrease and flow stream weakens. Velocity components with the increase in emissivity have not an appreciable variation for $Ra = 10^6$ based on Fig. 5b. It is confirmed in Fig. 6 that increase in emissivity also leads to decrease in temperature of the heat source surface for $Ra = 10^3$ and $Ra = 10^6$.

The effect of Rayleigh number on the dimensionless temperature field and streamlines is indicated in Fig. 7. It can be seen that isotherms have an elliptic shape in $Ra = 10^4$ which implies that diffusion is the dominated mode of heat transfer. The dominated mode of heat transfer is changed from diffusion to convection with the increase of Rayleigh and elliptic shape of isotherms is lost. As the Rayleigh number increases, isotherms layering lose the symmetry. In addition, as Rayleigh number increases, the temperature field decreases.

The effect of surface radiation on the dimensionless isotherms and streamlines is indicated in Fig. 8. Generally, it can be seen that the increase of emissivity decreases the temperature of heat source wall. Consequently temperature of more area of enclosure decreases. So mean temperature of enclosure decreases. Isotherms are also inclined at the coinciding to the walls with the increase of emissivity. It seems that the increase of emissivity strengthens diffusion mode of heat transfer as the streamlines get elliptic shape more.

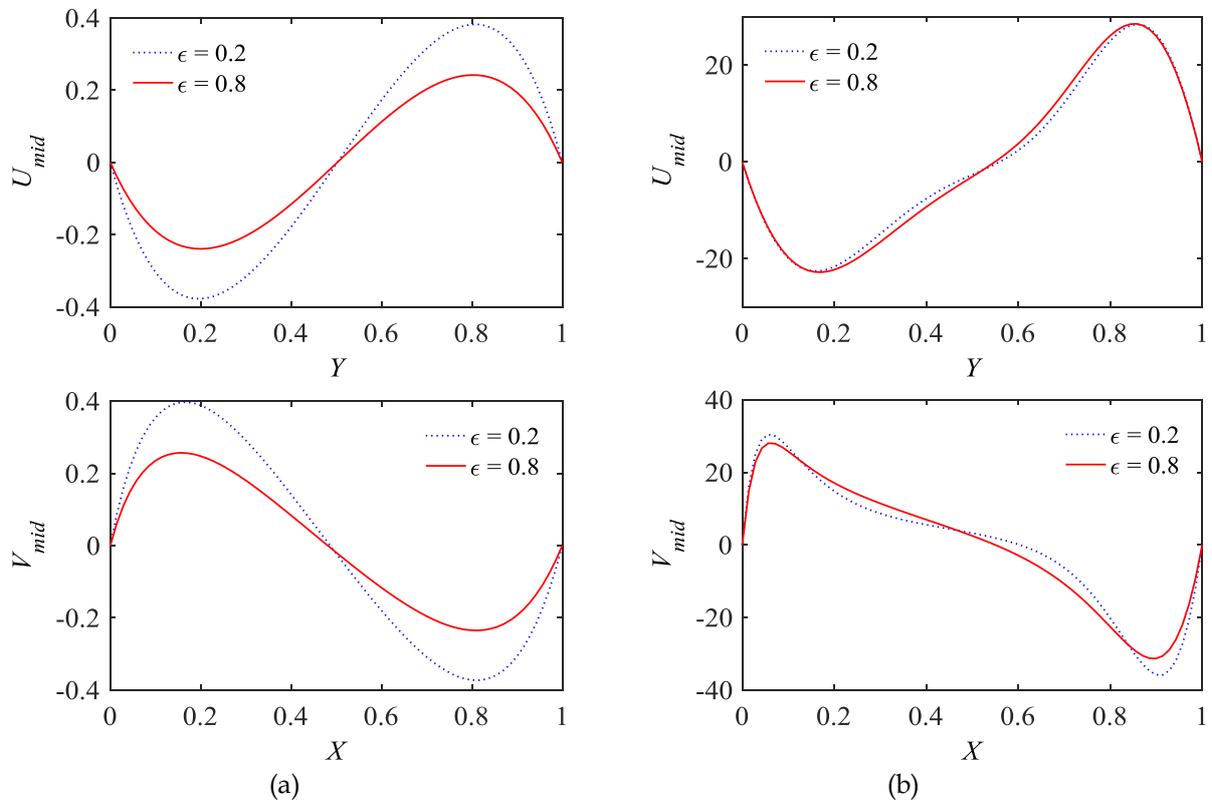

Fig. 5. Dimensionless velocity horizontal component profiles at plane $X = 0.5$ and dimensionless velocity vertical component profile at plane $Y = 0.5$ for $Y_h = 0.45, D_h = 0.1$ (a) $\text{Ra} = 10^3$ and (b) $\text{Ra} = 10^6$

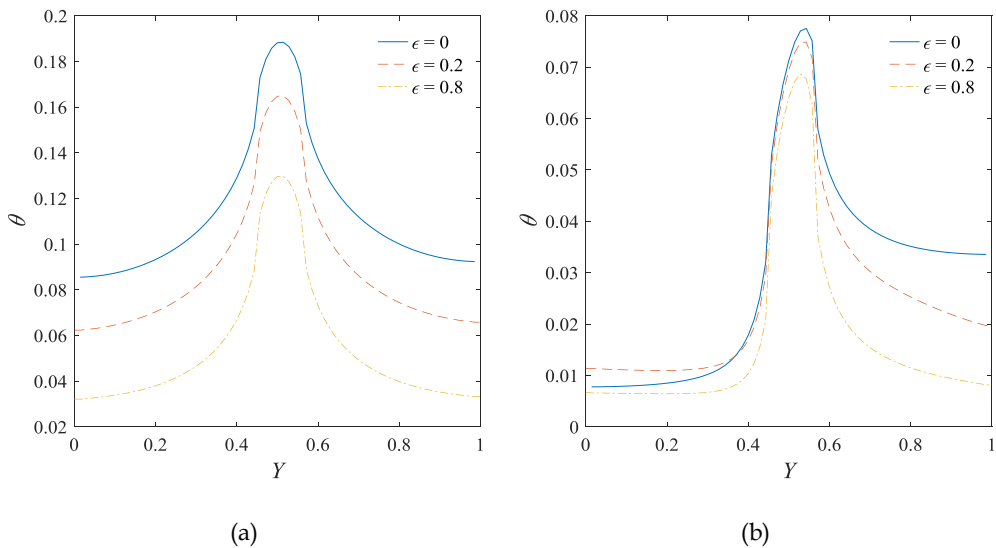

Fig. 6. Effect of surface radiation on the dimensionless temperature of heat source wall surface for $Y_h = 0.45, D_h = 0.1$ (a). $\text{Ra} = 10^3$ (b). $\text{Ra} = 10^6$

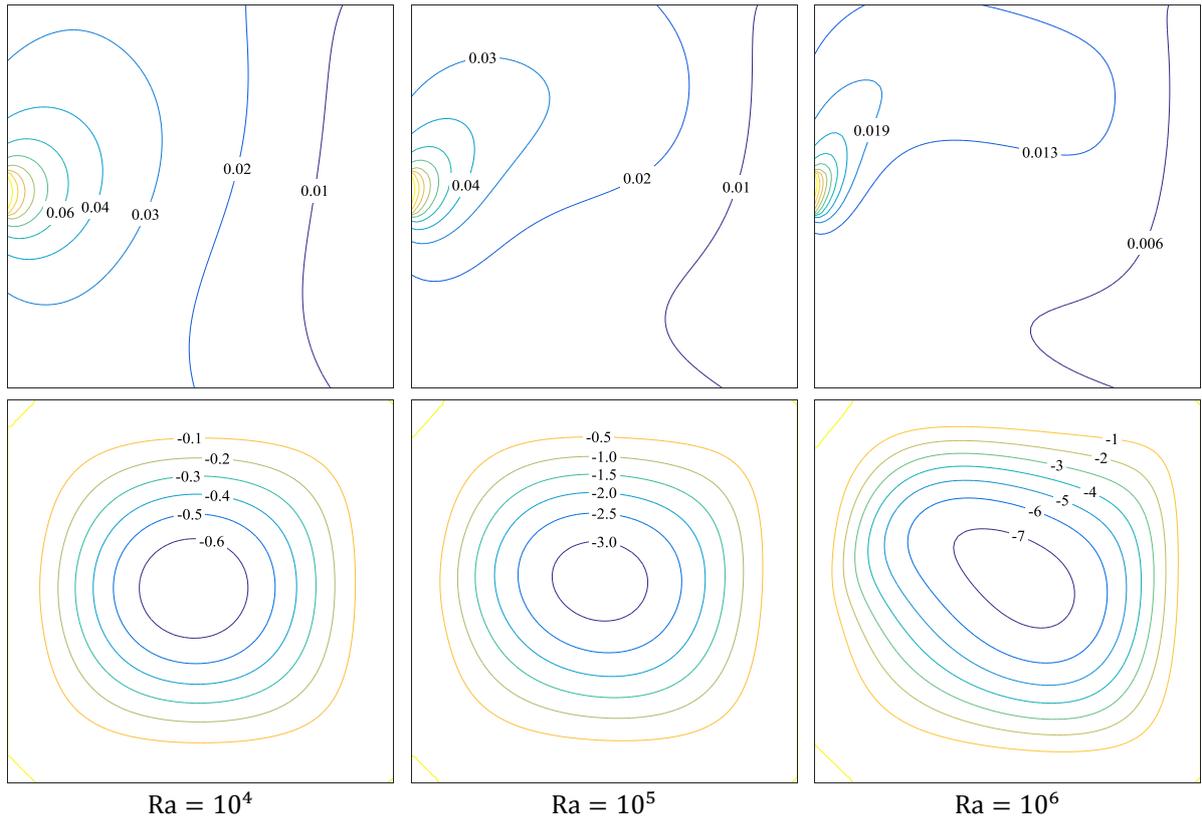

Fig. 7. Effect of Rayleigh number on the dimensionless temperature field (first row) and streamlines (second row) for $\varepsilon = 0.6$, $Y_h = 0.45$ and $D_h = 0.1$

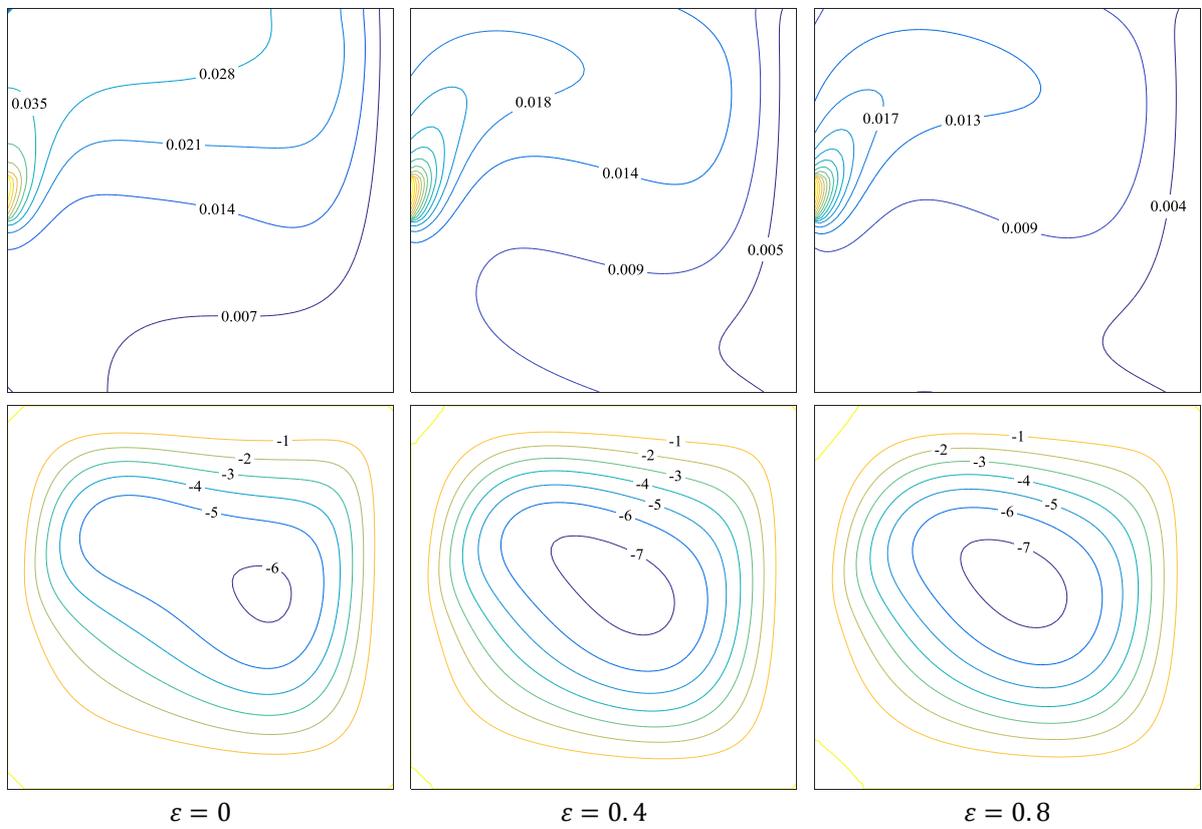

Fig. 8. Effect of emissivity on the dimensionless temperature field and streamlines for $Ra = 10^6$, $Y_h = 0.45$ and $D_h = 0.1$

## 4. 3. Effect of Rayleigh number and emissivity on the entropy generation

The effect of Rayleigh number on the dimensionless volumetric rate of entropy generation isolines due to the heat transfer ($N'''_{s,heat}$) and friction ($N'''_{s,fric}$) is shown in Fig. 9. It is seen that generally, as the

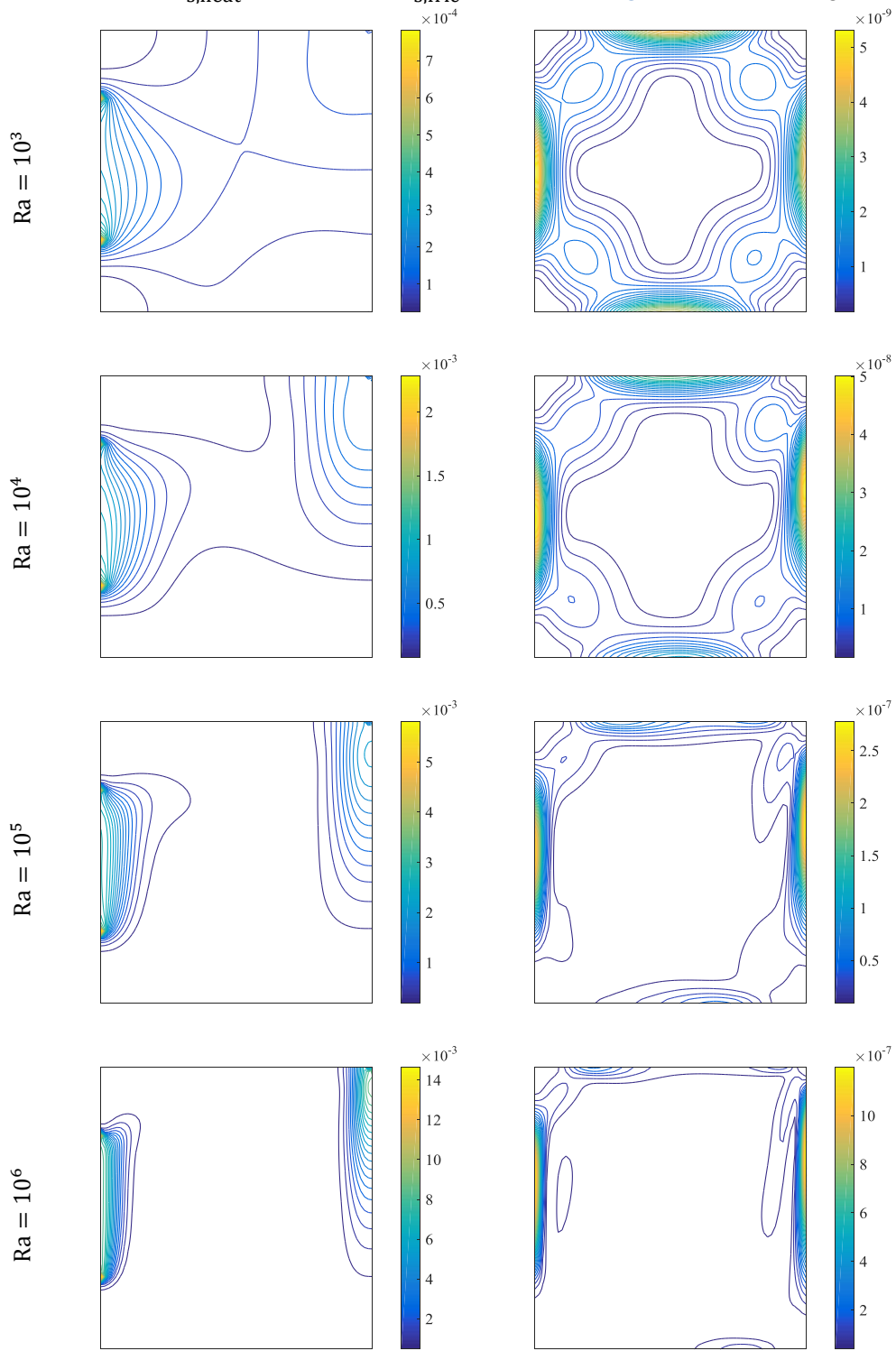

Fig. 9. Dimensionless volumetric rate of entropy generation isolines due to heat transfer (left column) and friction for (right column) different Rayleigh numbers ($\varepsilon = 0, Y_h = 0.25, D_h = 0.5$)

Rayleigh number increases, the rate of entropy generation increases. Mukhopadhyay [19] also showed this result. That's because of activation in stream movement and subsequently increase in heat transfer and friction losses occurs. The maximum rate of entropy generation happens due to the heat transfer happens near the both sides of the heat source that the maximum temperature gradients exist. With the increase in Rayleigh number, similar the temperature contours, the rate of entropy generation contours for heat transfer diminish the elliptic shape due to the convection domination mode per the diffusion. The fluid friction on the enclosure sides makes the entropy generation higher at these regions; however the left and right sides which stream is intensified, the entropy generation gets the larger values.

For a more concentration on the effect of Rayleigh number, the dimensionless volumetric rate of entropy generation profiles due to heat transfer and friction at crossing midplanes of the enclosure are presented in Fig. 10. Moreover the previous results, it is shown that the order of magnitude for the heat transfer losses is 4 times larger than the viscous dissipation losses; so the friction contribution in total rate of entropy generation is not efficacious.

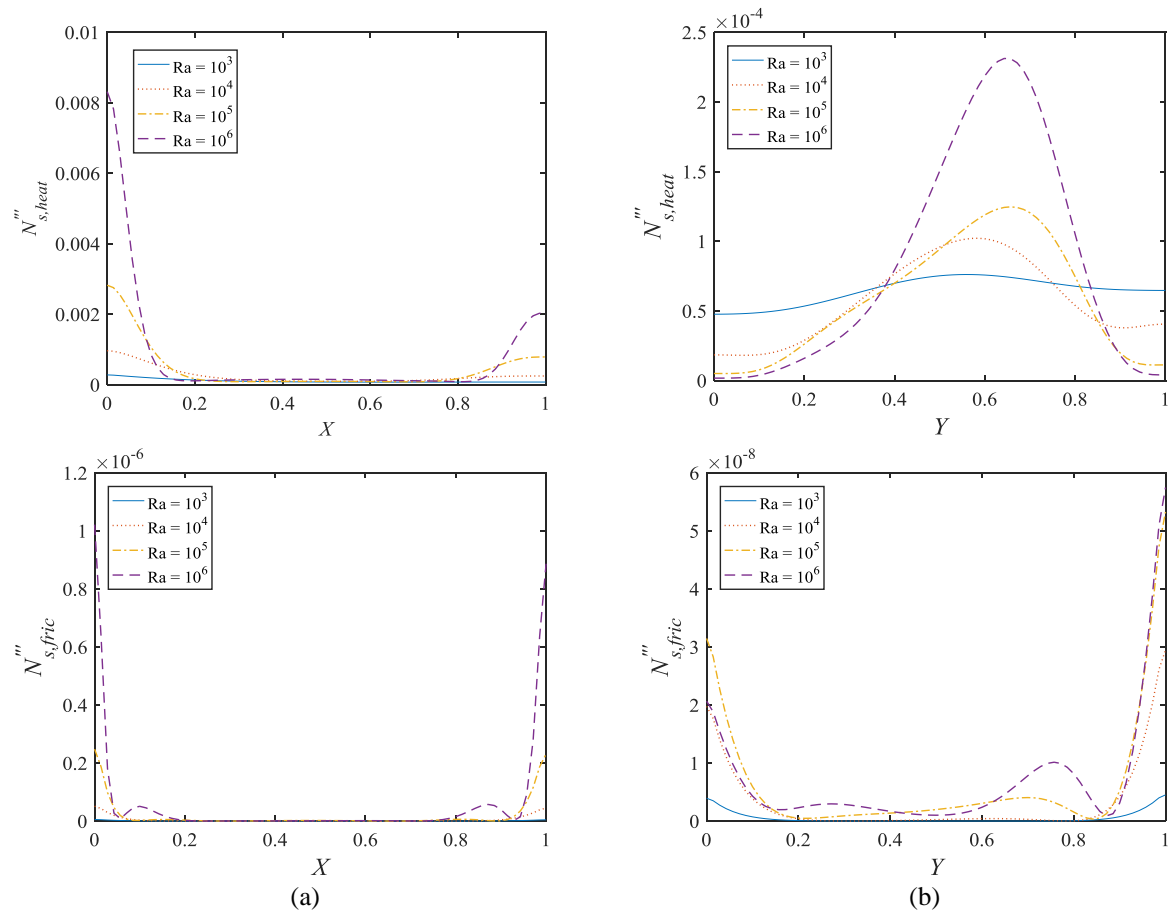

Fig. 10. Dimensionless volumetric rate of entropy generation profiles due to the heat transfer and friction at midplane (a). $Y = 0.5$ and (b). $X = 0.5$ in different Rayleigh numbers ( $\varepsilon = 0, Y_h = 0.25, D_h = 0.5$ )

Fig. 11 shows the effect of emissivity on the dimensionless volumetric rate of entropy generation profiles due to heat transfer and friction at crossing midplanes of the enclosure. As the emissivity increases, both entropy generations due to the heat transfer losses and friction decreases. It's due to the reduction in temperature which causes to decreases the losses. Hinojosa et al. [20] showed that for an open cavity with an isothermal heat source, the total dimensionless rate of entropy generation increases with the increase of emissivity; that is a difference with isoflux heat source.

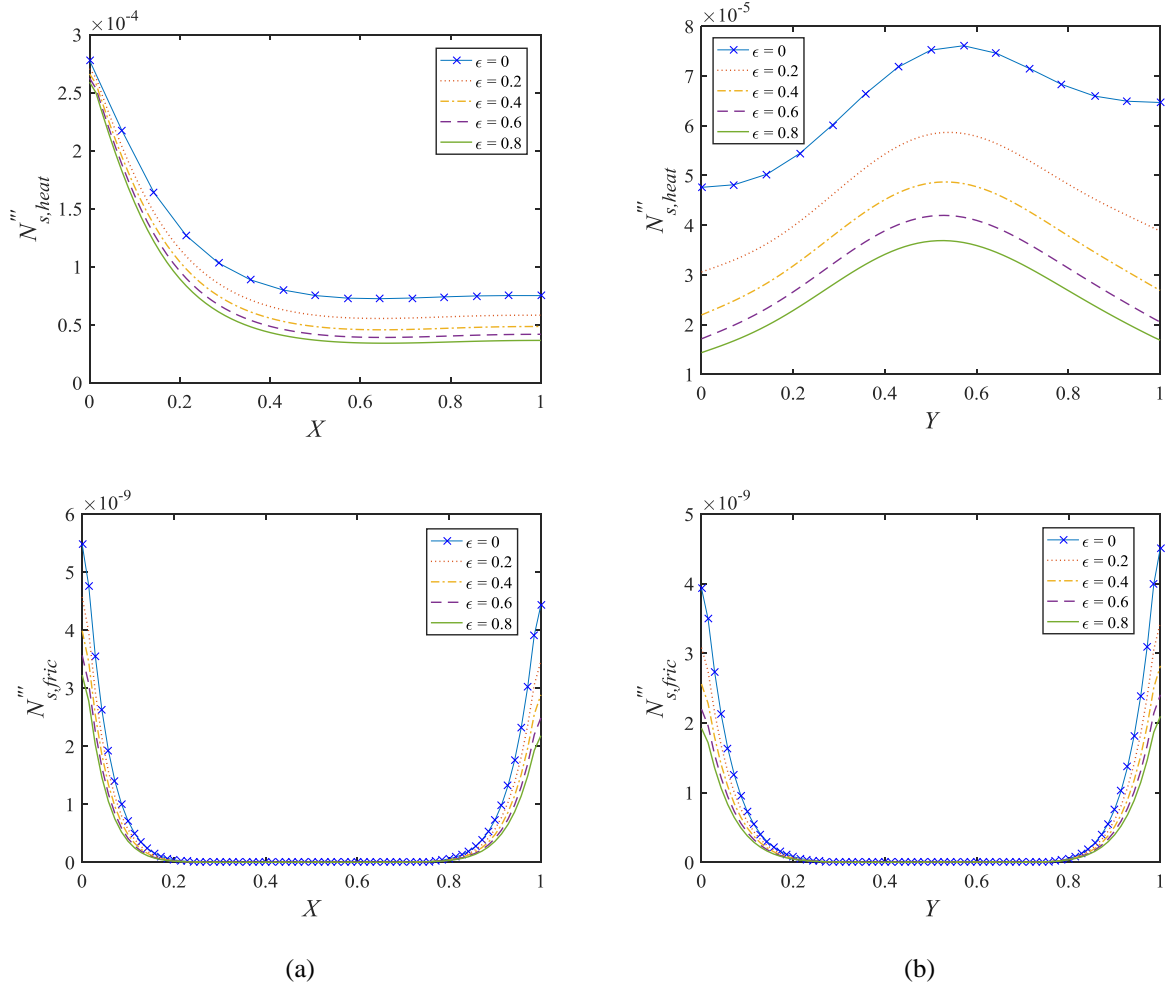

(a)                            (b)

Fig. 11. Dimensionless volumetric rate of entropy generation profiles due to the heat transfer and friction at midplane (a). $Y = 0.5$ and (b). $X = 0.5$ in different emissivity (Ra = $10^3, Y_h = 0.25, D_h = 0.5$)

A quantitative comparison between dimensionless average rate of entropy generation for system ($N_{s,\text{sys}}$) and dimensionless average rate of entropy generation due to the heat transfer and friction ($N_s$) at pure natural convection has been done in Table 3. The dimensionless average rate of entropy generation for system ($N_{s,sys}$) and total contributions due to the heat transfer and friction ($N_s$) conform to each other in pure natural convection state which it validates the results. It is also clear that all rates of entropy generation enhance with the increase in Rayleigh number.

Table 3 Dimensionless average rate of entropy generation for system ($N_{s,\text{sys}}$) and dimensionless average rate of entropy generation due to heat transfer and friction ($N_s$) at different Rayleigh numbers and pure natural convection $Y_h = 0.25$, $\varepsilon = 0$, $D_h = 0.5$

| Ra | H | $N_{s,\text{sys}}$ | $N_s$ | $N_{s,\text{heat}}$ | $N_{s,\text{fric}}$ |
|---|---|---|---|---|---|
| $10^3$ | 0.0123 | 7.7430×10⁻⁵ | 7.9421×10⁻⁵ | 7.9420×10⁻⁵ | 7.9852×10⁻¹⁰ |
| $10^4$ | 0.0219 | 1.5748×10⁻⁴ | 1.6138×10⁻⁴ | 1.6137×10⁻⁴ | 5.5230×10⁻⁹ |
| $10^5$ | 0.0389 | 2.8304×10⁻⁴ | 2.8936×10⁻⁴ | 2.8935×10⁻⁴ | 1.5805×10⁻⁸ |
| $10^6$ | 0.0691 | 5.2372×10⁻⁴ | 5.3173×10⁻⁴ | 5.3169×10⁻⁴ | 4.4358×10⁻⁸ |

**4.4. Optimal search of heat source location with EGM approach**

In EGM section, to achieve the optimum location of the heat source for all mentioned heat transfer parameters, the size of the heat source is kept fixed ($D_h = 0.1$). The effect of surface radiation on the dimensionless average rate of entropy generation for system ($N_{s,sys}$) with location $Y_h$ in different Rayleigh regimes is shown in Fig. 12. With the increase of Rayleigh in each emissivity, the dimensionless average rate of entropy generation for system increases. Since with the increase of Rayleigh, although the dimensionless temperature of the enclosure decreases (Fig. 6) and based on Eq. (26) heat source temperature reduction should lead to decrease the rate of entropy generation; but the height of the enclosure is extended while other parameters in Rayleigh number are fixed; finally this lead to increase the dimensionless average rate of entropy generation for system. It was also shown in previous section that the increase in Rayleigh number causes to increase in volumetric rate of entropy generation (Fig. 9 and Fig. 10) which composes a main part of entropy generation for system. With the increase of emissivity in each Rayleigh, the dimensionless average rate of entropy generation reduces. In addition to the reasons discussed in previous section (Fig. 11), it can be stated that the increase of emissivity causes to increase radiative heat flux from surfaces and subsequently decreases the temperature of the enclosure; hence the dimensionless average rate of entropy generation for system reduces and it adapts with Eq. (26). For $Ra = 10^3$ and $Ra = 10^4$, the optimum location of dimensionless average rate of entropy generation for $\varepsilon = 0$ to $\varepsilon = 0.8$ remains in center approximately ($Y_h = 0.45$). For $Ra = 10^5$, the variation of optimum location for similar routine is located in $Y_h = 0.4$ to $Y_h = 0.3$ approximately. For $Ra = 10^6$, by considering surface radiation up to $\varepsilon = 0.8$, the variation of optimum location for similar routine is located in $Y_h = 0.25$ to $Y_h = 0.35$ approximately.

As it was discussed before, other heat transfer parameters also can be as the optimal design criteria. Fig. 12 also indicates the effects of surface radiation on the maximum dimensionless temperature of the heat source in different positions. Generally, it is found that the minimum value of the maximum dimensionless temperature decreases with the increase of emissivity. As the Rayleigh number increases, the maximum dimensionless temperature decreases in the same emissivity, for each position. With increasing the Rayleigh number, the optimal position also approaches from the center to the bottom. There is a similarity among the curve variation of the dimensionless average rate of entropy generation and the maximum dimensionless temperature in each Rayleigh number and emissivity; the trend curve of the optimum location for the dimensionless average rate of entropy generation and the maximum dimensionless temperature is also close to each other. It can be said by using the study of maximum dimensionless temperature in the Fig. 12, the second law of thermodynamics is confirmed. This means that the optimal location of the heat source for dimensionless average rate of entropy generation in which irreversibility of the system is minimized, the maximum dimensionless temperature of the heat source surface is also minimized; hence, the EGM approach as the optimum design criteria is applicable. For $Ra = 10^3$ and $Ra = 10^4$, the optimum location of maximum dimensionless temperature by the change from $\varepsilon = 0$ to $\varepsilon = 0.8$ doesn't change considerably and remains at the center ($Y_h = 0.45$) approximately. For $Ra = 10^5$ and $Ra = 10^6$, the variation of optimum location for similar routine is located approximately in the interval of $0.3 \leq Y_h \leq 0.4$ and $0.25 \leq Y_h \leq 0.35$, respectively.

Another important heat transfer parameter which can be used as the optimal design criteria is the average convective Nusselt number that is a target to be maximized. The effect of surface radiation on the average convective and radiative Nusselt number with the location of the heat source in different Rayleigh numbers is shown in Fig. 13. As the heat source moves toward the around of the center of the wall, the average convective Nusselt number increases and then decreases till reach to the top. It is also an attention key that the increase of emissivity and consequently the increase of radiative heat transfer flux increases the average convective Nusselt number on the constant heat flux source unlike the observations for an isothermal heat source [3, 5-7, 9]. That is why although convective heat transfer flux decreases, temperature reduction is as highly that leads to the average convective Nusselt number increases based on Eq. (42). Fig. 6 also confirms that the increase of the emissivity leads to decrease the temperature of the heat source surface for $Ra = 10^3$ and $Ra = 10^6$. Based on Fig. 13, with the increase of Rayleigh for natural convection ($\varepsilon = 0$), the location of the maximum average convective Nusselt number shifts to the bottom of the enclosure and not at the center of the wall as it has resulted by Chu et al. [12] for the effect of isothermal heat source location in natural convection. Furthermore, Da Silva et al. [15] also found that the optimal location of constant heat flux source for the maximum of global conductance in natural convection shifts toward the bottom. Regarding Fig. 13, the location of maximum average convective Nusselt number for $Ra = 10^3$ and $Ra = 10^4$ in different emissivity occurs at the center ($Y_h = 0.45$) and has not appreciable change while the value of the maximum average convective Nusselt number varies significantly. For $Ra = 10^5$ and $Ra = 10^6$, by a change from $\varepsilon = 0$ to $\varepsilon = 0.8$, the optimum location changes in the interval of $0.3 \leq Y_h \leq 0.4$ and $0.25 \leq Y_h \leq 0.35$, respectively. There are some nuances between the optimal location for the average convective Nusselt number and dimensionless average rate of entropy generation; but there is still a good agreement between optimal location displacement trend.

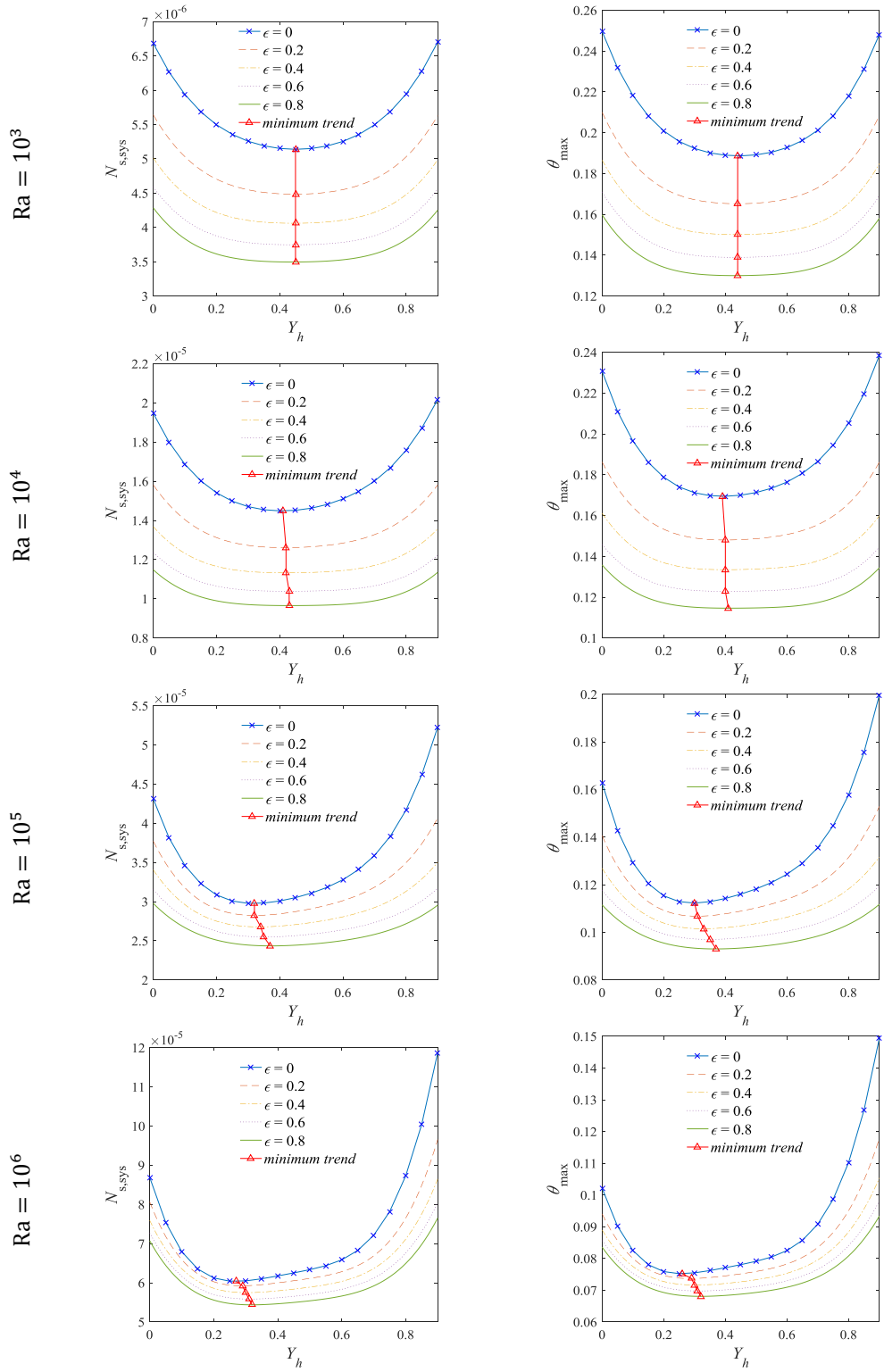

Fig. 12. Effect of surface radiation on the variation of the dimensionless average rate of entropy generation for system and maximum dimensionless temperature with location $Y_h$

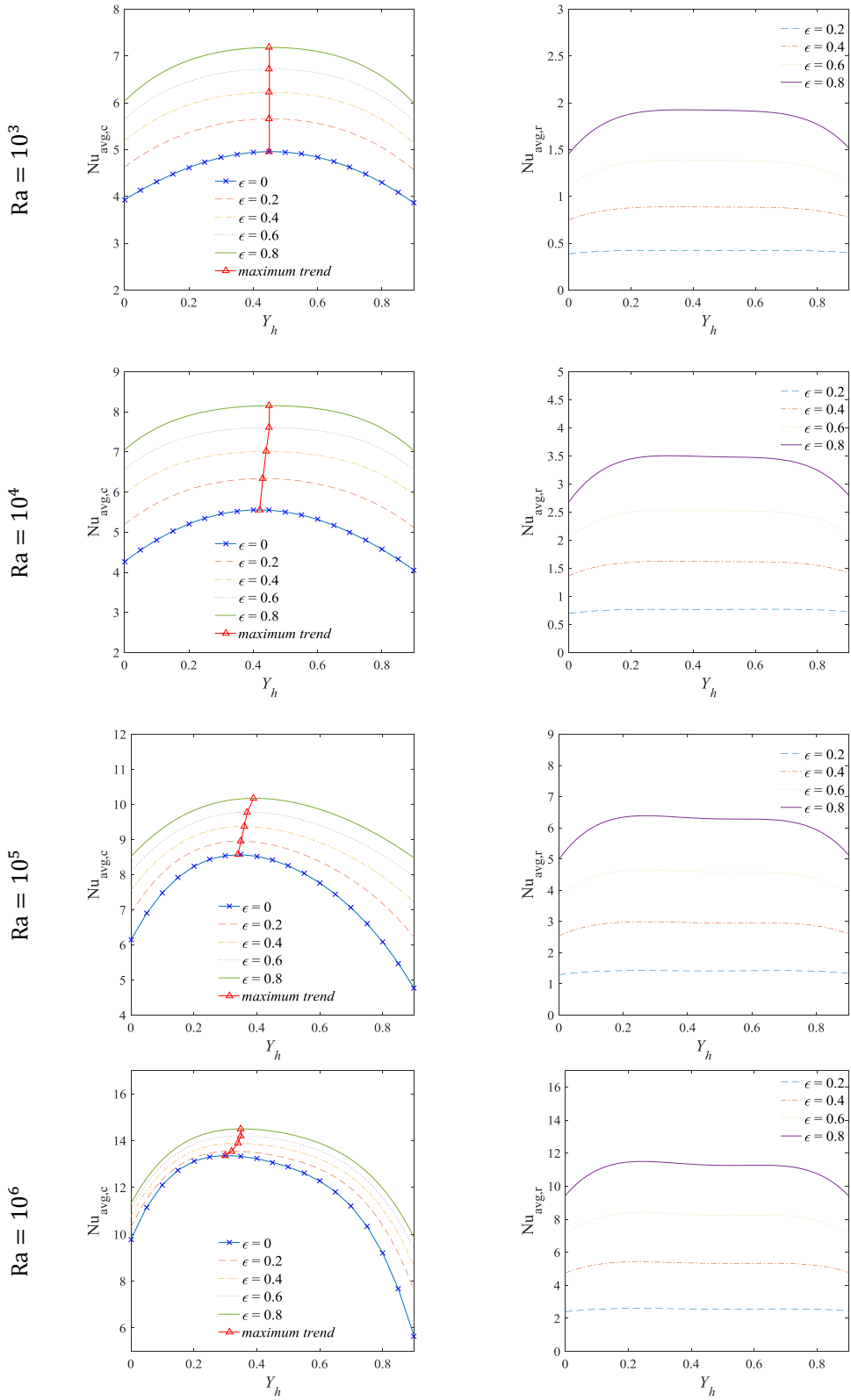

Fig. 13. Effect of surface radiation on the variation of the average convective and radiative Nusselt number with location $Y_h$

Based on Fig. 13 for the effect of surface radiation on the average radiative Nusselt number, it can be seen that generally, with the increase of Rayleigh, the average radiative Nusselt number increases. Furthermore, the increase of emissivity causes to increase in the average radiative Nusselt number in each location. Unlike the observations for the average convective Nusselt number, the average radiative Nusselt number in a certain emissivity and Rayleigh has not been affected in a wide range of location around the center of the wall approximately; So, the maximum average radiative Nusselt number is located in the center. For lower Rayleigh regimes ($Ra = 10^3, Ra = 10^4$), the maximum average radiative Nusselt number for variation $\varepsilon = 0.2$ to $\varepsilon = 0.8$ increases from 0.4 to 1.9 and 0.75 to 3.5, respectively. For higher Rayleigh regimes ($Ra = 10^5, Ra = 10^6$) the maximum average radiative Nusselt number for variation $\varepsilon = 0.2$ to $\varepsilon = 0.8$ also increases from 1.4 to 6.3 and 2.5 to 11.3, respectively. It can be concluded that considering surface radiation could multiply the average radiative Nusselt number for $\varepsilon \geq 0.2$ and it has a direct and linear relation with emissivity approximately.

## 5. Conclusions

In this study, analysis of conjugate natural convection with surface radiation in a two-dimensional enclosure was evaluated for an optimum design with EGM approach. Results were compared with pure natural convection and related previous researches were verified. In order to reach to an optimal design of system, dimensionless average rate of entropy generation is selected as the objective to be minimized. The other heat transfer parameters including maximum dimensionless temperature, average convective and radiative Nusselt numbers variations also followed with optimal search approach. The major conclusion remarks are as follows:

- By increase of emissivity, the temperature of heat source wall decreases. As the emissivity increases, in addition to both entropy generations due to the heat transfer losses and friction, entropy generation for system also decreases. As the Rayleigh number increases, the rate of entropy generation increases.

- By considering surface radiation up to $\varepsilon = 0.8$, the optimum dimensionless average rate of entropy generation could decrease 10% and the optimum location could alter 5% in higher Rayleigh numbers ($Ra = 10^5$, $Ra = 10^6$). In lower Rayleigh numbers ($Ra = 10^3$, $Ra =$ $10^4$), the optimum location of the dimensionless average rate of entropy generation had not considerable displacement; however, the optimum dimensionless average rate of entropy generation decreased up to 30%.

- Optimum design with other heat transfer parameters such as the maximum dimensionless temperature and convective Nusselt number confirmed the applicability of second law for optimal design with some slight differences due to numerical calculations. This means that the optimal location of the heat source for the minimum rate of entropy generation in which irreversibility of the system is minimized, the maximum dimensionless temperature of the heat source surface and convective Nusselt number are also optimized.